\documentclass[preprint,superscriptaddress,nofootinbib]{revtex4-1}
\usepackage{xcolor}
\usepackage{tabu}
\usepackage{titlesec}
\usepackage{graphicx,epsfig,epstopdf}
\usepackage{bm}
\usepackage{epstopdf}
\usepackage{amssymb,amsmath,amsfonts,latexsym}
\usepackage{dcolumn}
\usepackage{bm}
\usepackage{hyperref}
\usepackage[utf8]{inputenc}
\newcommand{\bea}{\begin{eqnarray}}
\newcommand{\ena}{\end{eqnarray}}
\newcommand{\be}{\begin{equation}}
\newcommand{\en}{\end{equation}}
\newcommand{\nn}{\nonumber\\}
\newcommand{\no}{\nonumber}
\newcommand{\ed}{\end{document}}

\newcommand{\Vcb}{V_{\mathrm{cb}}}
\newcommand{\BsDs}{B_s \to D_s^{(*)}\ell^+\nu_\ell}
\newcommand{\Ds}{D_s^{(*)}}
\usepackage{breqn}             
\makeatletter                  
\let\cat@comma@active\@empty   
\makeatother                   

\begin{document}
\title{Form factors and branching fraction calculations for $B_s \to D_s^{(*)} \ell^+ \nu_\ell$ in view of LHCb observation}
\author{N. R. Soni}
\email{nrsoni-apphy@msubaroda.ac.in}
\affiliation{Department of Physics, Faculty of Science, \\
The Maharaja Sayajirao University of Baroda, Vadodara 390002, Gujarat, {\it INDIA}}

\author{A. Issadykov}
\email{issadykov@jinr.ru}
\affiliation{Bogoliubov Laboratory of Theoretical Physics, \\
Joint Institute for Nuclear Research, 141980 Dubna, \textit{RUSSIA}}
\affiliation{The Institute of Nuclear Physics, \\
Ministry of Energy of the Republic of Kazakhstan, 050032 Almaty,  {\it KAZAKHSTAN}}
\affiliation{Al-Farabi Kazakh National University, 050040 Almaty, {\it KAZAKHSTAN}}

\author{A. N. Gadaria}
\affiliation{Applied Physics Department, Faculty of Technology and Engineering,  \\
The Maharaja Sayajirao University of Baroda, Vadodara 390001, Gujarat, {\it INDIA}}

\author{Z. Tyulemissov}
\affiliation{Bogoliubov Laboratory of Theoretical Physics, \\
Joint Institute for Nuclear Research, 141980 Dubna, \textit{RUSSIA}}
\affiliation{The Institute of Nuclear Physics, \\
Ministry of Energy of the Republic of Kazakhstan, 050032 Almaty,  {\it KAZAKHSTAN}}
\affiliation{Al-Farabi Kazakh National University, 050040 Almaty, {\it KAZAKHSTAN}}

\author{J. J. Patel}
\affiliation{Applied Physics Department, Faculty of Technology and Engineering, \\
The Maharaja Sayajirao University of Baroda, Vadodara 390001, Gujarat, {\it INDIA}}

\author{J. N. Pandya}
\email{jnpandya-phy@spuvvn.edu}
\affiliation{Department of Physics, Sardar Patel University, Vallabh Vidyanagar 388120, Gujarat, {\it INDIA}}

\date{\today}

\begin{abstract}
In light of the LHCb observations about the $B_s \to D_s^{(*)}\ell \nu_\ell$ semileptonic decays, we study these channels within the Standard Model framework of covariant confined quark model. The necessary transition form factors are computed in the entire dynamical range of momentum transfer squared with built-in infrared confinement.
Our computed ratios of the decay widths from tau mode to muon mode for $D_s$ and $D_s^*$ mesons are found to be $R(D_s) = 0.271 \pm 0.069$ and $R(D_s^*) = 0.240 \pm 0.038$. We further determine the ratio of the decay width from $D_s$ and $D_s^*$ channel for muon mode $\Gamma(B_s \to D_s \mu^+ \nu_\mu)/\Gamma(B_s \to D_s^* \mu^+ \nu_\mu) = 0.451 \pm 0.093$. Our results are in excellent agreement with the data from the latest LHCb experiments as well as lattice quantum chromodynamics simulations.
We also compare the shape of differential decay distribution for $B_s \to D_s^* \mu^+ \nu_\mu$ with the LHCb data and our results are in very good agreement throughout all the individual bins. Some other physical observables such as forward-backward asymmetry and longitudinal polarizations of leptons in the final state are also computed.
\end{abstract}

\maketitle

\section{Introduction}
\label{sec:introduction}
Semileptonic decays are one of the very important tool for studying the Physics of weak transitions. They have great phenomenological implications within and beyond the Standard Model (SM) of particle physics. Weak decays involve quark mixing and serve as a very important probe to measure the Cabibbo-Kobayashi-Maskawa (CKM) matrix elements.
Among other weak decays, the tree level transition $b \to c \ell \nu_\ell$ provides the coupling of heavy quark dynamics $\Vcb$\footnote{The inclusion of charge-conjugate processes is implied throughout this paper.}.
In the past few years, several new results have been reported by experimental facilities worldwide for $B \to D^{(*)} \ell \nu_\ell$ decays. Many of these results have violated the SM predictions suggesting the possibility of including new physics (NP) in these interactions. Lattice results are also available for studying these channels. These anomalies are reported in the most recent review article \cite{Albrecht:2021tul} and references therein.
$B_s \to D_s^{(*)}\ell \nu_\ell$ decay channels are very much similar to that of $B \to D^{(*)} \ell \nu_\ell$ decays except for the spectator quark and therefore similar anomalies are also expected from these channels.
Further, it can also be employed as a potential candidate for the determination of  $V_{\mathrm{cb}}$ and also the search for NP beyond the standard model.
LHCb collaboration has measured the $|\Vcb|$ from transition form factors and decay rates of $B_s^0 \to D_s^- \mu^+\nu_\mu$ and $B_s^0 \to D_s^{*-} \mu^+\nu_\mu$ channels for the very first time \cite{Aaij:2020hsi}.
They have also measured the absolute branching fractions for the first time and their results are found to be \cite{Aaij:2020hsi}
\begin{eqnarray}
\mathcal{B} (B_s^0 \to  D_s^- \mu^+\nu_\mu)_{\mathrm{LHCb}} &=&  (2.49 \pm 0.12 (\mathrm{stat}) \pm 0.14 (\mathrm{syst}) \pm 0.16 (\mathrm{ext})) \times 10^{-2},\nn
\mathcal{B} (B_s^0 \to  D_s^{*-} \mu^+\nu_\mu)_{\mathrm{LHCb}} & = & (5.38 \pm 0.25 (\mathrm{stat}) \pm 0.46 (\mathrm{syst}) \pm 0.30 (\mathrm{ext})) \times 10^{-2}.
\end{eqnarray}
They also determined the ratio of both the channels and it is found to be
\begin{eqnarray}
\frac{\mathcal{B} (B_s^0 \to D_s^- \mu^+\nu_\mu)_{\mathrm{LHCb}}}{\mathcal{B} (B_s^0 \to D_s^{*-} \mu^+\nu_\mu)_{\mathrm{LHCb}}} &=& 0.464 \pm 0.013 (\mathrm{stat}) \pm 0.043 (\mathrm{syst}).
\end{eqnarray}

Precise Lattice results on the determination of $\Vcb$ are also available as the higher valence $s$ quark mass in $B_s$ transition form factors compared to light quark masses in $B$ mesons making it comparatively less intensive for computation.
Recently, HPQCD collaboration has provided the transition form factors in the entire $q^2$ range for the channels $B_s \to D_s$ \cite{McLean:2019qcx} and $B_s \to D_s^*$ \cite{Harrison:2021tol}. They have also computed the semileptonic branching fractions and the ratio of the branching fractions reads
\begin{eqnarray}
\frac{\mathcal{B} (B_s^0 \to D_s^- \mu^+\nu_\mu)_{\mathrm{HPQCD}}}{\mathcal{B} (B_s^0 \to D_s^{*-} \mu^+\nu_\mu)_{\mathrm{HPQCD}}} &=& 0.429~(43)_{\mathrm{latt}}~(4)_{\mathrm{EM}},
\end{eqnarray}
which is in very good agreement with the LHCb data.
Further, they have also provided the ratios $R(D_s)$ and $R(D_s^*)$ which is the ratio of branching fractions of tau mode to electron or muon mode as \cite{McLean:2019qcx,Harrison:2021tol}
\begin{eqnarray}
R(D_s)_{\mathrm{HPQCD}} &=& \frac{\mathcal{B}(B_s \to D_s \tau \nu_\tau)}{\mathcal{B}(B_s \to D_s \ell \nu_\ell)} = 0.2993~(46), \nn
R(D_s^*)_{\mathrm{HPQCD}} &=& \frac{\mathcal{B}(B_s \to D_s^* \tau \nu_\tau)}{\mathcal{B}(B_s \to D_s^* \ell \nu_\ell)} = 0.2442~(79)_{\mathrm{latt}} (35)_{\mathrm{EM}}. 
\end{eqnarray}
In literature, the transition form factors and semileptonic branching fractions for $B_s \to D_s^{(*)}\ell \nu_\ell$ decays have been reported within the SM framework of perturbative QCD factorization approach with lattice inputs of transition form factors \cite{Hu:2019bdf}.
Very recently,  $B_s \to D_s$ transition form factors and semileptonic branching fractions were also computed in leading as well as higher twist distribution amplitude using light cone QCD sum rules \cite{Zhang:2021wnv,  Bordone:2019guc, Li:2009wq}.
Exclusive semileptonic decays were also reported in the framework of three point sum rules calculation \cite{Blasi:1993fi}.
Different quark models have also studied the transition form factors and semileptonic branching fractions viz. light front quark model \cite{Verma:2011yw,Kang:2018jzg,Li:2010bb}, relativistic quark model \cite{Faustov:2012mt}, instantaneous Bethe–Salpeter equation \cite{Chen:2011ut}, constituent quark model \cite{Zhao:2006at}. 

In this article, we provide a detailed study for semileptonic decays of $B_s \to D_s^{(*)} \ell \nu_\ell$ channels where $\ell = e, \mu, \tau$. The necessary transition form factors are computed in the entire momentum transfer squared range $0 < q^2 < (m_{B_s} - m_{D_s^{(*)}})^2$ within the quantum field theoretical framework of covariant confined quark model (CCQM) developed by G. V. Efimov and M. A. Ivanov \cite{Efimov:1993,Branz:2009cd,Ivanov:2011aa,Gutsche:2012ze,Soni:2020bvu}.
We then compute the semileptonic branching fractions without using any additional parameters.
Further, as a probe to lepton flavor universality, we also determine the ratio of branching fractions for electron or muon to tau mode for both $D_s$ and $D_s^*$ mesons.
With the advancement of worldwide experimental facilities including CERN, we expect data regarding physical observables such as forward-backward asymmetries, different polarising observables and others in near future. Hence, we have computed these physical observables in the present study.
We compare our findings with the LHCb, lattice QCD data and other theoretical predictions.
In recent years, we have been successful in employing CCQM for the study of semileptonic decays of charmed mesons \cite{Soni:2017eug,Soni:2018adu,Ivanov:2019nqd,Soni:2020sgn,Soni:2019huk,Soni:2019qjs,Soni:2021rem,Dubnicka:2017job,Issadykov:2018jvt,Issadykov:2017wlb}.
Further, Ivanov \textit{et. al.,} have also computed the $B \to D^{(*)}\ell\nu_\ell$ decay channels in the Refs \cite{Ivanov:2015tru,Ivanov:2016qtw,Ivanov:2017mrj}.
Together with Refs. \cite{Ivanov:2015tru,Ivanov:2016qtw,Ivanov:2017mrj}, this work complements the theoretical study of the semileptonic decays for the transition $b \to c \ell \nu_\ell$ and thus tests the validity of CCQM for mesonic transitions.

This paper is organized as follows.
After a brief introduction along with the literature survey, we introduce essential  components of CCQM and provide the formulation for computing the branching fractions and other physical observables in Sec. \ref{sec:framework}. We also provide the model parameters and transition form factors in the entire dynamical range of momentum transfer.
Next, in Sec. \ref{sec:results}, we present results for the decay $\mathcal{B}(B_s \to D_s^{(*)}\ell^+\nu_\ell)$ for $\ell = e, \mu$ and $\tau$.
We compare our findings with the experimental data, lattice simulation and other theoretical predictions.
We further compute the $q^2$ averages of forward-backward asymmetry and longitudinal polarization.
Finally, in Sec. \ref{sec:conclusion}, we summarise and conclude the present work.
\section{Theoretical Framework}
\label{sec:framework}
Within the standard model framework, and neglecting QED corrections,  the semileptonic transition form factors for $B_s$ decays to $\Ds$ mesons can be written as
\begin{eqnarray}
\mathcal{M} (\BsDs) = \frac{G_F}{\sqrt{2}} \Vcb \langle \Ds | \bar{c} O^\mu b|B_s \rangle [\ell^+ O_\mu \nu_\ell],
\end{eqnarray}
where $G_F$ is the Fermi coupling constant and $O^\mu = \gamma^\mu (1 - \gamma_5)$ is the Dirac matrix with the left chirality.
The matrix element in this decay channel is well parameterized in terms of Lorentz invariant form factors in terms of momentum transferred squared between parent ($B_s$) and daughter ($\Ds$) mesons are given by
\begin{eqnarray}
\langle D_s (p_2) &|& \bar{c} O^\mu b | B_s (p_1) \rangle  =  F_+ (q^2) P^\mu + F_- (q^2) q^\mu , \nn
\langle D_s^* (p_2, \epsilon_\nu) &|& \bar{c} O^\mu b | B_s (p_1) \rangle  =  \frac{\epsilon_{\nu}^{\dag}}{m_1 + m_2} \left[ -g^{\mu\nu} P\cdot q A_0(q^2)  \right. \cr && \left. + P^{\mu} P^{\nu}  A_+(q^2) + q^{\mu} P^{\nu}  A_-(q^2) +  i\varepsilon^{\mu\nu\alpha\beta} P_{\alpha} q_{\beta}V(q^2) \right ],
\label{eq:form_factor_general}
\end{eqnarray}
with $P = p_1 + p_2$,  $q = p_1 - p_2$ and $\epsilon_{\nu}$ to be the polarization vector such that $\epsilon_{\nu}^{\dag} \cdot p_2 = 0$ and on-shell conditions of particles require $p_1^2 = m_1^2 = m_{B_s}^2$ and $p_2^2 = m_2^2 = m_{D_s^{(*)}}^2$.
With the help of form factors, the semileptonic branching fractions can be computed using the model-independent formulation.
The semileptonic differential decay rates corresponding to the transition $b \to c \ell^+ \nu_\ell$ can be written in terms of the helicity amplitudes can be written as
\begin{eqnarray}
\frac{d\Gamma(\BsDs)}{dq^2} = \frac{G_F^2 |V_{cb}|^2 |{\bf p_2}| q^2}{96 \pi^3m_1^2}  (1 - 2 \delta_\ell)^2  \times  \Big[(1+ \delta_\ell) |H_n|^2 + 3 \delta_\ell |H_t|^2\Big].
\label{eq:decay_width}
\end{eqnarray}
Here, $|H_n|^2 = |H_+|^2 + |H_-|^2 + |H_0|^2$, $|{\bf p_2}| = \lambda^{1/2} (m_1^2, m_2^2, q^2)/2 m_1$ is the momentum of the daughter meson in the rest frame of parent meson with $\lambda$ is the K\"allen function and $\delta_\ell = m_\ell^2/2q^2$ is the helicity flip factor.
In the above Eq. (\ref{eq:decay_width}), the bilinear combinations of the helicity amplitudes are defined in terms of form factors for the channel $B_s \to D_s \ell^+ \nu_\ell$ as
\begin{eqnarray}
\nonumber
H_t &=& \frac{1}{\sqrt{q^2}} (Pq F_+ + q^2 F_-),\\
H_\pm &=& 0 \ \ \ {\textrm and} \ \ \ H_0 = \frac{2 m_1 |\bf{p_2}|}{\sqrt{q^2}} F_+,
\end{eqnarray}
and for the channel $B_s \to D_s^* \ell^+ \nu_\ell$
\begin{eqnarray}
\nonumber
H_t  &=& \frac{1}{m_1+m_2} \frac{m_1 |{\bf p_2}|}{m_2\sqrt{q^2}} \left((m_1^2 - m_2^2) (A_+ - A_-) + q^2 A_-\right),\\
H_\pm &=& \frac{1}{m_1+m_2} (-(m_1^2 - m_2^2)A_0 \pm 2 m_1 |{\bf p_2}| V), \nonumber \\
H_0 &=& \frac{1}{m_1+m_2} \frac{1}{2m_2\sqrt{q^2}} (-(m_1^2 - m_2^2)(m_1^2 - m_2^2-q^2)A_0  + 4m_1^2 |{\bf p_2}|^2A_+).
\end{eqnarray}
Several other physical observables including forward-backward asymmetry $\mathcal{A}_{FB}^\ell(q^2)$ and longitudinal $P_L^\ell(q^2)$ polarization of the leptons in terms of helicity amplitudes, in the final state, can be written as
\begin{eqnarray}
\mathcal{A}_{FB}^\ell(q^2) &=&
-\frac{3}{4}
\frac{|H_+|^2 - |H_-|^2 + 4 \delta_\ell H_0 H_t}{(1 + \delta_\ell) |H_n|^2 + 3\delta_\ell |H_t|^2}, \nn
P_L^\ell(q^2) &=&
-\frac{(1 - \delta_\ell) |H_n|^2 - 3\delta_\ell |H_t|^2}{(1 + \delta_\ell) |H_n|^2 + 3\delta_\ell |H_t|^2}.
\label{eq:observables}
\end{eqnarray}
Note that to compute the averages of these observables, one has to multiply the numerator and denominator by the phase-space factor $|{\bf p_2}| q^2 (1 - m_{\ell}^2/q^2)^2$ and integrate separately.

Semileptonic transition form factors appearing in Eq. (\ref{eq:form_factor_general}) are computed within the SM framework of CCQM which is the effective quantum field theoretical approach for hadronic interaction with the constituent quarks  \cite{Efimov:1993,Branz:2009cd,Ivanov:2011aa,Gutsche:2012ze,Soni:2020bvu}.
For hadronic system, the interaction Lagrangian describing the coupling of a meson with 	constituents is written as
\begin{eqnarray}
\mathcal{L}_{\mathrm{int}}  &=& g_M M(x) \int dx_1\int dx_2 F_M(x;x_1,x_2) \bar{q}_2(x_2) \Gamma_M q_1(x_1) + \mathrm{H.c}.
\label{eq:lagrangian}
\end{eqnarray}
Here $M(x)$ is the hadron field of investigated meson. The Dirac matrix $\Gamma_M = \gamma_5, \gamma_\mu$ are used for pseudoscalar and vector mesons, respectively.
$g_M$ is the coupling strength for meson with constituents which can be determined using the renormalization of self-energy diagram employing compositeness condition \cite{Weinberg:1962,Salam:1962}. 
This condition essentially guarantees that the final mesonic state does not contain any bare quark state and also avoids the double-counting of the hadronic degree of freedom.
In above Eq. (\ref{eq:lagrangian}), $F_M(x;x_1,x_2)$ is the vertex function of the four dimensional coordinate that describes the quark distribution within the meson and hence characterizes the physical size of the meson given by
\begin{eqnarray}
F_M(x;x_1,x_2) = \delta \left(x - \sum_{i=1}^2 w_i x_i \right) \Phi_M \big((x_1 - x_2)^2\big),
\label{eq:vertex}
\end{eqnarray}
with $w_i = m_{q_i}/(m_{q_1} + m_{q_2})$.
$\Phi_M$ is the correlation function of the constituent quarks and further to avoid any divergences in the Feynman diagram, the Fourier transform of $\Phi_M$ must have an appropriate falloff behaviour in the Euclidean region. For computation, we choose the vertex function to be the Gaussian function of the form
\begin{eqnarray}
\tilde{\Phi}_M (-p^2) = \mathrm{exp} (p^2/\Lambda_M^2).
\label{eq:Gaussian}
\end{eqnarray}
Note that different other forms for the vertex functions are also employed, however it is observed that hadronic observables are independent of the detailed structure of vertex function \cite{Ivanov:1997ug,Faessler:2003yf}. 
Here, the model parameter $\Lambda_M$ is characterized by the finite size of the meson.
Matrix elements for the self-energy diagram and any other decay process are described by the convolution of propagators and vertex functions. For semileptonic decay $\BsDs$, the matrix element can be written in the Minkowski space as
\begin{eqnarray}
\langle D_s (p_2) | \bar{x}O^\mu c  | B_s(p_1) \rangle  &=&
N_c g_{B_s} g_{D_s} \int \frac{d^4 k}{(2\pi)^4 i}  \tilde{\Phi}_{B_s} (-(k + w_{13} p_1)^2) \tilde{\Phi}_{D_s}(-(k + w_{23} p_2)^2) \nonumber \\ &\times &  \mathrm{tr}[O^\mu S_1(k + p_1) \gamma^5 S_3(k) \gamma^5 S_2(k + p_2)], \nn
\langle D_s^* (p_2,\epsilon_{\nu}) | \bar{x}O^\mu c  | B_s(p_1) \rangle  &=& N_c g_{B_s} g_{D_s^*} \int \frac{d^4 k}{(2\pi)^4 i} \tilde{\Phi}_{B_s} (-(k + w_{13} p_1)^2) \tilde{\Phi}_{D_s^*}(-(k + w_{23} p_2)^2) \nonumber \\ &\times &  \mathrm{tr}[O^\mu S_1(k + p_1) \gamma^5 S_3(k) \not\!{\epsilon}_{\nu}^\dag S_2(k + p_2)].
\label{eq:ff}
\end{eqnarray}
\begin{figure*}[htbp]
\includegraphics[width=0.45\textwidth]{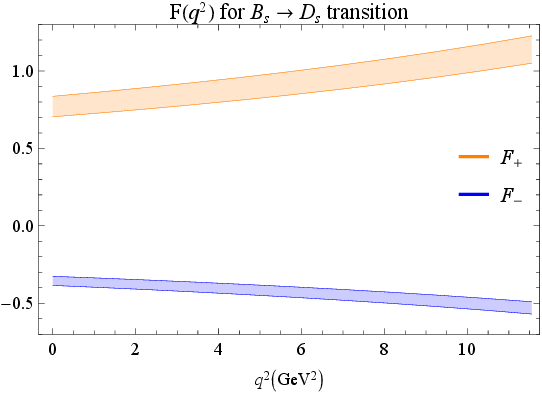}
\hfill\includegraphics[width=0.45\textwidth]{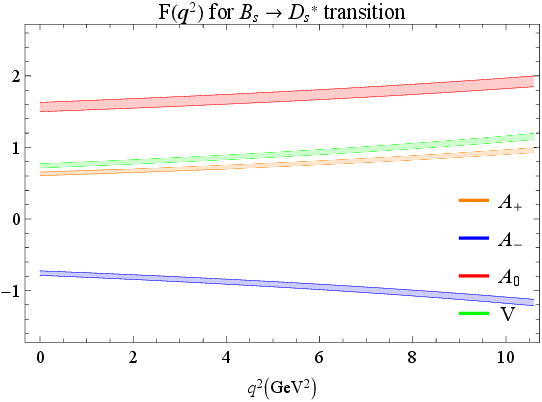}
\caption{Form factors}
\label{fig:form_factors}
\end{figure*}
Here $N_c = 3$ is the number of colors and $w_{ij} = m_{q_i}/(m_{q_i} + m_{q_j})$. $S_{1,2,3}$ are the quark propagator and here we use the Fock-Schwinger representation for the propagator. This allows us to do the loop integration in efficient way as it transforms the loop momenta to the derivative of the external momenta.
Finally, to remove any divergence in the quark loop diagram, $\lambda$ an infrared cutoff parameter is introduced which essentially guarantees the quark confinement within a hadron. 
We consider $\lambda = 0.181$ GeV to be universal for all hadronic interactions studied using CCQM \cite{Branz:2009cd}.
A detailed description of CCQM including the computation techniques  can be found in the Ref. \cite{Branz:2009cd,Ivanov:2019nqd}.
The model parameters such as quark masses $m_q$ and meson size parameter $\Lambda_M$ are determined for the basic electromagnetic properties like leptonic decay constants \cite{Ivanov:2011aa}.
For present computations, we employ model parameters that are obtained using updated least square fit procedure performed in the Refs.  \cite{Ivanov:2019nqd,Soni:2020sgn,Ivanov:2015tru}. 
These parameters are fitted with the available experimental results and lattice simulations of the weak  leptonic decay constants of $B_s$ and $D_s^{(*)}$ mesons.
Further,  the parametrization was achieved to keep the deviation in the computed decay constants defined by the function 
\begin{eqnarray}
\chi^2 =\sum_i \frac{(y_i^{\textrm{experiment}} - y_i^{\textrm{theory}})^2}{\sigma_i^2} \no 
\end{eqnarray}
to be minimum \cite{Ivanov:2016qtw,Ivanov:2017mrj}. 
Here, $\sigma_i$ are reported experimental standard deviations. 
After all the size parameters were fitted to get the best possible decay constant values, the uncertainties in the respective parameters were determined by individually changing them to get the exact experimental or lattice results. The difference between these two values of the parameters were considered as uncertainty in the respective parameter. We have checked for all flavoured mesons and this uncertainty is found to be within 5\% for all the parameters. These uncertainties are considered absolute for given parameters and are then transported to the form factors in the whole $q^2$ range and it is observed that the uncertainties in the form factors are less than 10 \% at the maximum recoil.
The spread of the uncertainties in the form factors is displayed in Fig. \ref{fig:form_factors}.
The form factors appearing in Eq. (\ref{eq:form_factor_general}) are also very well represented in terms of double pole approximation
\begin{eqnarray}
F(q^2) = \frac{F(0)}{1 - a \left(\frac{q^2}{m_{B_s}^2}\right) + b \left(\frac{q^2}{m_{B_s}^2}\right)^2}.
\label{eq:double_pole}
\end{eqnarray}
The form factor at maximum recoil $F(0)$ and associated double pole parameters are tabulated in Tab. \ref{tab:form_factor}.
For present computation, we use the model parameters such as quark masses $m_b = 5.05$ GeV, $m_c = 1.672$ GeV and $m_s = 0.428$ GeV and size parameters $\Lambda_{B_s} = 2.05 \pm 0.014$ GeV, $\Lambda_{D_s} = 1.75 \pm 0.035$ GeV and $\Lambda_{D_s^*} = 1.56 \pm 0.014$ GeV \cite{Ivanov:2019nqd,Soni:2020sgn,Ivanov:2015tru}.
Using these parameters, the coupling strengths have been computed using the compositeness condition and the coupling strengths are found to be $g_{B_s} = 4.960 \pm 0.111$,  $g_{D_s} = 3.813 \pm 0.197$ and $g_{D_s^*} = 2.544 \pm 0.066$. 
All other parameters such as meson masses, the lifetime of $B_s$ meson, CKM matrix element and Fermi coupling constant are sourced from the Particle Data Group \cite{ParticleDataGroup:2020ssz} and CKM matrix element $V_{\mathrm{cb}} = 0.04221$ \cite{Alberti:2014yda}.
\begin{table*}
\caption{Form factors and double pole parameters appeared in Eq. (\ref{eq:double_pole})}
\begin{tabular*}{\textwidth}{@{\extracolsep{\fill}}cccccccc@{}}
\hline\hline
$F$ & $F(0)$ & $a$ & $b$ & $F$ & $F(0)$ & $a$ & $b$\\
\hline
$F_+^{B_s \to D_s}$ 		& $0.770 \pm 0.066$ & 0.837 & 0.077 & $F_-^{B_s \to D_s}$ &  $-0.355 \pm 0.029$ & 0.855 & 0.083 \\
$A_+^{B_s \to D_s^*}$ 	& $0.630 \pm 0.025$ & 0.972 & 0.092 & $A_-^{B_s \to D_s^*}$ & $-0.756 \pm 0.031$ & 1.001 & 0.116\\
$A_0^{B_s \to D_s^*}$ 	& $1.564 \pm 0.065$ & 0.442 & $-$0.178 & $V^{B_s \to D_s^*}$ & $0.743 \pm 0.030$ & 1.010 & 0.118\\
\hline\hline
\label{tab:form_factor}
\end{tabular*}
\end{table*}
\begin{table*}
\caption{Comparison of the form factor at maximum recoil with relativistic quark model (RQM), perturbative QCD (PQCD), sum rules (SR) and light front quark model (LFQM).  }
\begin{tabular*}{\textwidth}{@{\extracolsep{\fill}}lccccc@{}}
\hline\hline
 & $F_{+} (0)$ & $V(0)$ & $A_0(0)$ & $A_1(0)$ & $A_2(0)$\\
\hline
Present & $0.770 \pm 0.066$ & $0.743 \pm 0.030$ & $0.719 \pm 0.070$ & $0.681 \pm 0.065$ & $0.630 \pm 0.025$\\
RQM \cite{Faustov:2012mt} & $0.74 \pm 0.02$ & $0.95 \pm 0.02$ & $0.67 \pm 0.01$ & $0.70 \pm 0.01$ & $0.75 \pm 0.02$\\
PQCD \cite{Hu:2019bdf} & $0.52 \pm 0.10$ & $0.64 \pm 0.12$ & $0.48 \pm 0.09$ & $0.50 \pm 0.09$ & $0.53 \pm 0.11$\\
SR \cite{Blasi:1993fi} & $0.7 \pm 0.1$ & 0.63 $\pm 0.05$ & $ 0.52 \pm 0.06$ & $0.62 \pm 0.1$ & $0.75 \pm 0.07$\\
LFQM \cite{Li:2010bb} & ... & $0.74^{+0.05}_{-0.05}$ & $0.63^{+0.04}_{-0.04}$ & $0.61^{+0.04}_{-0.04}$ & $0.59^{+0.04}_{-0.04}$\\
\hline\hline
\label{tab:form_factor_comparison}
\end{tabular*}
\end{table*}
\begin{figure*}[htbp]
\includegraphics[width=0.45\textwidth]{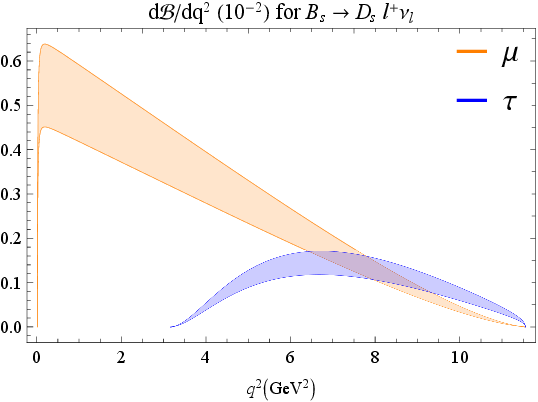}
\hfill\includegraphics[width=0.45\textwidth]{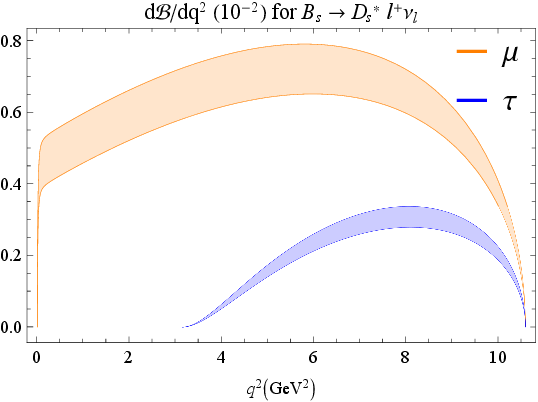}
\caption{Differential decay rates for the channels $B_s \to D_s^{(*)}\ell \nu_\ell$ for $\ell = \mu$ and $\tau$. }
\label{fig:decay}
\end{figure*}
\begin{figure}[htbp]
\centering
\includegraphics[width=0.45\textwidth]{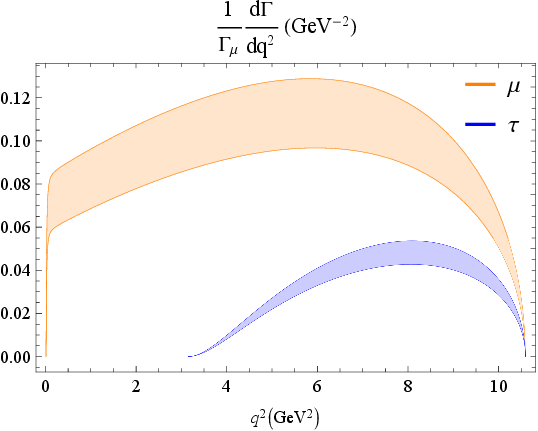}
\caption{Normalized differential decay rates for the channels $B_s \to D_s^*\ell \nu_\ell$ for $\ell = \mu$ and $\tau$. }
\label{fig:norm_decay}
\end{figure}
\section{Results and Discussion}
\label{sec:results}
\begin{figure}[htbp]
\centering
\includegraphics[width=0.45\textwidth]{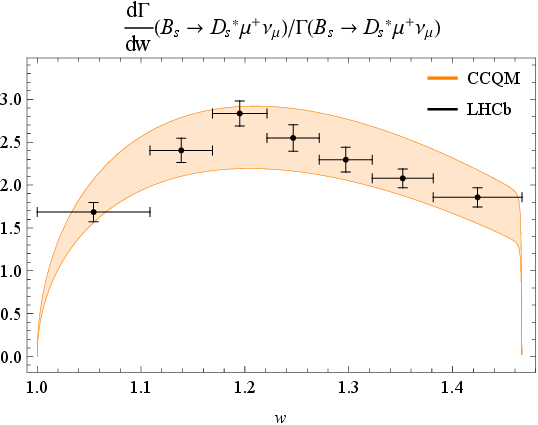}
\caption{Normalized differential decay rates for the channels $B_s \to D_s^* \mu^+ \nu_\mu$ as a function of recoil parameter Eq. (\ref{eq:recoil}) along with the comparison with LHCb data \cite{LHCb:2020hpv}.}
\label{fig:lhcb_decay}
\end{figure}
\begin{table*}
\caption{Normalized decay rates for channel $B_s \to D_s^* \mu^+ \nu_\mu$ in different recoil parameter $w$ bins along with the comparison with lattice QCD and LHCb data. }
\begin{tabular*}{\textwidth}{@{\extracolsep{\fill}}cccc@{}}
\hline\hline
$w$ bin & Present & LQCD \cite{Harrison:2021tol} & LHCb \cite{LHCb:2020hpv}\\
\hline
$1.0 - 1.1087$		& $0.183 \pm 0.019$ & 0.187 (11) 		& 0.183 (12)\\
$1.1087 - 1.1688$	& $0.146 \pm 0.015$ & 0.1507 (60) 	& 0.144 (84) \\
$1.1688 - 1.2212$	& $0.133 \pm 0.019$ & 0.1371 (38)	& 0.148 (76)\\
$1.2212 - 1.2717$	& $0.127 \pm 0.018$ & 0.1296 (24) 	& 0.128 (77)\\
$1.2717 - 1.3226$	& $0.123 \pm 0.018$ & 0.1230 (26) 	& 0.117 (69)\\
$1.3226 - 1.3814$	& $0.130 \pm 0.020$ & 0.1275 (54) 	& 0.122 (62)\\
$1.3814 - 1.4667$	& $0.157 \pm 0.026$ & 0.145 (15) 		& 0.158 (93)\\
\hline\hline
\end{tabular*}
\label{tab:bins}
\end{table*}
\begin{table*}
\caption{$B_s \to D_s^{(*)} \ell \nu_\ell$ Branching fractions (in \%)}
\begin{tabular*}{\textwidth}{@{\extracolsep{\fill}}lcccccc@{}}
\hline\hline
Channel & Present & RQM \cite{Faustov:2012mt} & LFQM \cite{Kang:2018jzg} & LFQM \cite{Li:2010bb} & PQCD \cite{Hu:2019bdf} & LHCb \cite{Aaij:2020hsi}\\
\hline
$B_s^0 \to D_s^- e^+ \nu_e$ 		& $2.89 \pm 0.50$ & 2.1 $\pm$ 0.2		& $2.45 \pm 0.27$	& ...  &   $1.84^{+0.77}_{-0.51}$	& ...   \\
$B_s^0 \to D_s^- \mu^+ \nu_\mu$ & $2.88 \pm 0.49$ & ...  & ...  & ...  & ...  &$2.49 \pm 0.12 \pm 0.14 \pm 0.16$ \\
$B_s^0 \to D_s^- \tau^+ \nu_\tau$ 	& $0.78 \pm 0.15$ & 0.62 $\pm$ 0.05	& $0.733 \pm 0.081$ & ...   &  $0.63^{+0.17}_{-0.13}$	& ...  \\
$B_s^0 \to D_s^{*-} e^+ \nu_e$ 	& $6.42 \pm 0.67$ & 5.3 $\pm$ 0.5		& $6.05 \pm 0.67$	&  ...  & $4.42^{+1.27}_{-1.00}$	\\
$B_s^0 \to D_s^{*-} \mu^+ \nu_\mu$ & $6.39 \pm 0.67$ & ...  & ...  & $5.2^{+0.6}_{-0.6}$ &  ...   & $5.38 \pm 0.25 \pm 0.46 \pm 0.30$ \\
$B_s^0 \to D_s^{*-} \tau^+ \nu_\tau$& $1.53 \pm 0.15$ & 1.3 $\pm$ 0.1		& $1.51 \pm 0.17$	& $1.3^{+0.2}_{-0.1}$ &  $1.20^{+0.26}_{-0.23}$	& ...  \\
\hline\hline
\end{tabular*}\label{tab:decay_width}
\end{table*}
\begin{table*}
\caption{Ratios of the semileptonic decay widths with heavy quark expansion (HQE), LHCb and  lattice QCD data}
\begin{tabular*}{\textwidth}{@{\extracolsep{\fill}}llll@{}}
\hline\hline
Ratio & $R(D_s) = \frac{\Gamma(B_s^0 \to D_s^- \tau^+ \nu_\tau)}{\Gamma(B_s^0 \to D_s^- \mu^+ \nu_\mu)}$ & $R(D_s^*) = \frac{\Gamma(B_s^0 \to D_s^{*-} \tau^+ \nu_\tau)}{\Gamma(B_s^0 \to D_s^{*-} \mu^+ \nu_\mu)}$ & $\frac{\Gamma(B_s^0 \to D_s^- \mu^+ \nu_\mu)}{\Gamma(B_s^0 \to D_s^{*-} \mu^+ \nu_\mu)}$\\
\hline
Present & $0.271 \pm 0.069$ 	& $0.240 \pm 0.034$ 	& $0.451 \pm 0.096$\\
HQE \cite{Bordone:2019guc}			& $0.2971 \pm 0.0034$ & $0.2472 \pm 0.0077$ & ...  \\
PQCD \cite{Hu:2019bdf} 		& $0.341^{+0.024}_{-0.025}$ & $0.271^{+0.015}_{-0.016}$ & ...  \\
LQCD \cite{McLean:2019qcx,Harrison:2021tol} 		& $0.2993 (46)$ & $0.2442 (79)_{\mathrm{latt}} (35)_{\mathrm{EM}}$ & $0.429 (43)_{\mathrm{latt}} (4)_{\mathrm{EM}}$ \\
LHCb \cite{Aaij:2020hsi} 		& ...  	& ...   & $0.464 \pm 0.013 \pm 0.043$\\
\hline\hline
\end{tabular*}
\label{tab:ratio}
\end{table*}
\begin{table*}
\caption{Averages of forward-backward asymmetry and longitudinal polarisation}
\begin{tabular*}{\textwidth}{@{\extracolsep{\fill}}ccccccc@{}}
\hline\hline
Channel & $\langle A_{FB}^e \rangle $ & $\langle A_{FB}^\mu \rangle $ & $\langle A_{FB}^\tau \rangle $ & $\langle P_L^e \rangle$ & $\langle P_L^\mu \rangle$ & $\langle P_L^\tau \rangle$\\
\hline
$B_s \to D_s$ 	 & $-1.16 \times 10^{-6}$	& $-$0.015	& $-$0.362	& $-$1.000	& $-$0.958	& 0.179\\
$B_s \to D_s^*$ & 0.195							& 0.190 		& 0.029			& $-$1.000	& $-$0.985	& $-$0.515\\
\hline\hline
\end{tabular*}
\label{tab:observables}
\end{table*}
Using the model parameters such as quark masses and size parameters, we first compute the transition form factors in the entire physical range of momentum transfer within the standard model framework of covariant confined quark model explained in the previous section.
In Tab. \ref{tab:form_factor_comparison},  we compare our form factors at maximum recoil with other studies such as relativistic quark model,  perturbative QCD,  light front quark model and QCD sum rules.  
It is observed that our results are in good agreement (within uncertainty) particularly with light front quark model \cite{Li:2010bb}.  Our results are also in good agreement with the relativistic quark model studies \cite{Faustov:2012mt} except for the form factor $V(0)$.  Also our results are systematically higher than perturbative QCD approach \cite{Hu:2019bdf}. 
Note that in order to compare our form factors of vector daughter meson Eq.  (\ref{eq:form_factor_general}) with the other approaches, we have transformed our form factors to Bauer-Stech-Wirbel form factors \cite{Wirbel:1985ji}. 
The transition form factors are then employed for the computation of semileptonic differential decay rates using Eq. (\ref{eq:decay_width}) and plotted in Fig. \ref{fig:decay}.
In Fig. \ref{fig:norm_decay}, we also plot the normalized differential decay distribution for the channels $B_s \to D_s^*\ell^+\nu_\ell$ for muon and tau mode. Note here that the normalization is achieved using the total decay rate for muon mode.
In these differential plots, the spread corresponds to the propagated uncertainty that arises solely from the form factors.
LHCb collaboration recently measured normalized decay distribution for the channel $B_s^0 \to D_s^{*-}\mu^+ \nu_\mu$ against the hadronic recoil parameter $w$ \cite{LHCb:2020hpv}.  Later on, HPQCD collaboration also computed the differential branching fractions and also provided the normalised differential decay rates in terms of hadronic recoil parameter \cite{Harrison:2021tol}. 
We also present our results in terms of $w$ to have the proper comparison with LHCb data and lattice simulation.
In the rest frame of $B_s$ meson, the momentum transfer squared $q^2$ is transformed in terms of the recoil parameter $w$ via the relation \cite{Harrison:2021tol}
\begin{eqnarray}
w = \frac{m_{B_s}^2 + m_{D_s^*}^2 - q^2}{2 m_{B_s} m_{D_s^*}}.
\label{eq:recoil}
\end{eqnarray}
In Fig. \ref{fig:lhcb_decay}, we plot the normalized differential decay distribution in terms of recoil parameters. We also show here the recent LHCb data \cite{Aaij:2020hsi} and it is observed that where our results are in excellent agreement with LHCb data.
Further, in Tab. \ref{tab:bins} we compute the normalized decay rates in small recoil parameter bins and also compare with lattice QCD and LHCb data. Our results are in excellent agreement with each of the bins as well.

We also compute the semileptonic branching fractions by integrating the differential branching fractions as shown in Fig. \ref{fig:decay} and compare with the different theoretical approaches along with LHCb data in Tab. \ref{tab:decay_width} and it is observed that our results are within the uncertainty presented LFQM results \cite{Kang:2018jzg}.
Previously also, our results were in very good agreement with the LFQM studies for the semileptonic $D_{(s)}$ studies.
For $\ell = e$ or $\mu$ mode, our branching fraction results overestimate the RQM \cite{Faustov:2012mt} and PQCD \cite{Hu:2019bdf} predictions, however for $\ell = \tau$, our results are within the uncertainties predicted by PQCD. The disagreement mainly arises due to the difference between our form factors and those used in RQM and PQCD.
For $B_s \to D_s$ channel too, our results are well within the range predicted by the constituent quark model \cite{Zhao:2006at}.
The reader may note that our results for $B_s^0 \to D_s^{(*)-}\mu^+ \nu_\mu$ are within the range predicted recently by LHCb collaboration \cite{Aaij:2020hsi}.
We also compute the ratios of the decay width of the tau channel to the muon channel and the results are given in Tab. \ref{tab:ratio} and it is observed that our results are in excellent agreement with recent lattice and LHCb data.
It is very important to note here that these ratios $R(D_s^{(*)})$ allow one to probe for the test of lepton flavor universality.
Present results are also in good agreement with the $R(D^{(*)})$ and $R(J/\psi)$ computed using this model in the Ref. \cite{Ivanov:2015tru,Issadykov:2018myx}.
The Belle collaboration has also measured the most precise measurement of ratios of decay widths and their results are found to be $R(D) = 0.307 \pm 0.037 \pm 0.016$ and $R(D^*) = 0.283 \pm 0.018 \pm 0.014$ \cite{Belle:2019rba}.
It is interesting to note that CCQM results on $R(D_{(s)}^{(*)})$ are well within the uncertainty predicted by Belle collaboration \cite{Belle:2019rba}. However, considering only the central value, our predictions are nearly 16\% lower then Belle data.
In Tab. \ref{tab:ratio} we also present the ratio of decay rates of $B_s \to D_s$ and $B_s \to D_s^*$ for muon channel and it is observed that our results are also in very good agreement with lattice and LHCb data.

Next, we compute other physical observables such as forward-backward asymmetry and longitudinal polarization using the relations Eq. (\ref{eq:observables}) and their $q^2$ average values are tabulated in Tab. \ref{tab:observables}.
These physical observables are identified experimentally for the other semileptonic decays at $B$ factories, however, they are yet to be reported for the channels considered here. These observables are also extremely sensitive and useful for testing the lepton flavor violating decays and thus can serve as probes for the physics beyond the standard model.
\section{Summary and Conclusion}
\label{sec:conclusion}
In this article,  in view of LHCb results, we have provided a detailed study of the semileptonic decay $B_s \to D_s^{(*)} \ell^+\nu_\ell$ for $\ell = e, \mu$ and $\tau$ in the framework of the covariant confined quark model. The in-built feature of infrared confinement helps us to remove divergences in the quark loop diagrams.
We have computed the transition form factors in the entire dynamical range of momentum transfer squared and employed them for the determination of semileptonic differential decay rates.
We have tested our model at individual $q^2$ for comparing with the lattice simulation and very  recent experimental data using the normalized differential decay distribution for the channels $B_s \to D_s^* \ell^+ \nu_\ell$.
Our calculation of the normalized decay distribution as a function of recoil parameters is in very good agreement with the LHCb data and HPQCD predictions. Further, our computed normalized decay rates are also in excellent agreement with them within reported uncertainties.
Our results of the semileptonic branching fractions are consistent with the LHCb data, lattice QCD simulation, perturbative QCD approach with lattice inputs, relativistic quark model and light-front quark model within reported uncertainties.
The ratios of the decay widths are also in very good agreement with the experimental data and other literature.
Present CCQM results on $R(D_s^{(*)})$ are consistent with recent experimental data by Belle collaboration within their uncertainties suggesting no violation of lepton flavor universality.
Other physical observables such as forward-backward asymmetries and longitudinal polarization are also computed in the present study.
The LHCb results are yet to be reported for the absolute semileptonic branching fractions for the tau mode for the transition $B_s \to \Ds$. Also, other $B$ factories are yet to explore precisely these exclusive transitions.
Exclusive semileptonic decays of $B_s$ are more advantageous compared to that of $B$ meson as background pollution from partially reconstructed decays is less severe than $B$ mesons and their counterparts. We anticipate more detailed results from these facilities in very shorty.

\section{Data Availability Statement}
There is no data associated with the manuscript.

\section*{ACKNOWLEDGEMENTS}
We would like to thank Prof. Mikhail A. Ivanov for useful discussions of some aspects of this work.
J.N.P. acknowledges financial support from University Grants Commission of India under Major Research Project F.No. 42-775/2013(SR), DST-PURSE, DST-FIST and UGC-DRS schemes.
N.R.S. would also like to thank Akash Hingu for his help in plotting Fig. \ref{fig:lhcb_decay}.
This research has been funded by the Science Committee of the Ministry of Education and Science of the Republic of Kazakhstan (Grant No. AP09057862).

\bibliography{apssamp}

\begin{thebibliography}{43}%
\makeatletter
\providecommand \@ifxundefined [1]{%
 \@ifx{#1\undefined}
}%
\providecommand \@ifnum [1]{%
 \ifnum #1\expandafter \@firstoftwo
 \else \expandafter \@secondoftwo
 \fi
}%
\providecommand \@ifx [1]{%
 \ifx #1\expandafter \@firstoftwo
 \else \expandafter \@secondoftwo
 \fi
}%
\providecommand \natexlab [1]{#1}%
\providecommand \enquote  [1]{``#1''}%
\providecommand \bibnamefont  [1]{#1}%
\providecommand \bibfnamefont [1]{#1}%
\providecommand \citenamefont [1]{#1}%
\providecommand \href@noop [0]{\@secondoftwo}%
\providecommand \href [0]{\begingroup \@sanitize@url \@href}%
\providecommand \@href[1]{\@@startlink{#1}\@@href}%
\providecommand \@@href[1]{\endgroup#1\@@endlink}%
\providecommand \@sanitize@url [0]{\catcode `\\12\catcode `\$12\catcode
  `\&12\catcode `\#12\catcode `\^12\catcode `\_12\catcode `\%12\relax}%
\providecommand \@@startlink[1]{}%
\providecommand \@@endlink[0]{}%
\providecommand \url  [0]{\begingroup\@sanitize@url \@url }%
\providecommand \@url [1]{\endgroup\@href {#1}{\urlprefix }}%
\providecommand \urlprefix  [0]{URL }%
\providecommand \Eprint [0]{\href }%
\providecommand \doibase [0]{http://dx.doi.org/}%
\providecommand \selectlanguage [0]{\@gobble}%
\providecommand \bibinfo  [0]{\@secondoftwo}%
\providecommand \bibfield  [0]{\@secondoftwo}%
\providecommand \translation [1]{[#1]}%
\providecommand \BibitemOpen [0]{}%
\providecommand \bibitemStop [0]{}%
\providecommand \bibitemNoStop [0]{.\EOS\space}%
\providecommand \EOS [0]{\spacefactor3000\relax}%
\providecommand \BibitemShut  [1]{\csname bibitem#1\endcsname}%
\let\auto@bib@innerbib\@empty
\bibitem [{\citenamefont {Albrecht}\ \emph {et~al.}(2021)\citenamefont
  {Albrecht}, \citenamefont {van Dyk},\ and\ \citenamefont
  {Langenbruch}}]{Albrecht:2021tul}%
  \BibitemOpen
  \bibfield  {author} {\bibinfo {author} {\bibfnamefont {J.}~\bibnamefont
  {Albrecht}}, \bibinfo {author} {\bibfnamefont {D.}~\bibnamefont {van Dyk}}, \
  and\ \bibinfo {author} {\bibfnamefont {C.}~\bibnamefont {Langenbruch}},\
  }\href {\doibase 10.1016/j.ppnp.2021.103885} {\bibfield  {journal} {\bibinfo
  {journal} {Prog. Part. Nucl. Phys.}\ }\textbf {\bibinfo {volume} {120}},\
  \bibinfo {pages} {103885} (\bibinfo {year} {2021})},\ \Eprint
  {http://arxiv.org/abs/2107.04822} {arXiv:2107.04822 [hep-ex]} \BibitemShut
  {NoStop}%
\bibitem [{\citenamefont {Aaij}\ \emph
  {et~al.}(2020{\natexlab{a}})\citenamefont {Aaij} \emph
  {et~al.}}]{Aaij:2020hsi}%
  \BibitemOpen
  \bibfield  {author} {\bibinfo {author} {\bibfnamefont {R.}~\bibnamefont
  {Aaij}} \emph {et~al.} (\bibinfo {collaboration} {LHCb Collaboration}),\
  }\href {\doibase 10.1103/PhysRevD.101.072004} {\bibfield  {journal} {\bibinfo
   {journal} {Phys. Rev. D}\ }\textbf {\bibinfo {volume} {101}},\ \bibinfo
  {pages} {072004} (\bibinfo {year} {2020}{\natexlab{a}})},\ \Eprint
  {http://arxiv.org/abs/2001.03225} {arXiv:2001.03225 [hep-ex]} \BibitemShut
  {NoStop}%
\bibitem [{\citenamefont {McLean}\ \emph {et~al.}(2020)\citenamefont {McLean},
  \citenamefont {Davies}, \citenamefont {Koponen},\ and\ \citenamefont
  {Lytle}}]{McLean:2019qcx}%
  \BibitemOpen
  \bibfield  {author} {\bibinfo {author} {\bibfnamefont {E.}~\bibnamefont
  {McLean}}, \bibinfo {author} {\bibfnamefont {C.~T.~H.}\ \bibnamefont
  {Davies}}, \bibinfo {author} {\bibfnamefont {J.}~\bibnamefont {Koponen}}, \
  and\ \bibinfo {author} {\bibfnamefont {A.~T.}\ \bibnamefont {Lytle}},\ }\href
  {\doibase 10.1103/PhysRevD.101.074513} {\bibfield  {journal} {\bibinfo
  {journal} {Phys. Rev. D}\ }\textbf {\bibinfo {volume} {101}},\ \bibinfo
  {pages} {074513} (\bibinfo {year} {2020})},\ \Eprint
  {http://arxiv.org/abs/1906.00701} {arXiv:1906.00701 [hep-lat]} \BibitemShut
  {NoStop}%
\bibitem [{\citenamefont {Harrison}\ and\ \citenamefont
  {Davies}(2022)}]{Harrison:2021tol}%
  \BibitemOpen
  \bibfield  {author} {\bibinfo {author} {\bibfnamefont {J.}~\bibnamefont
  {Harrison}}\ and\ \bibinfo {author} {\bibfnamefont {C.~T.~H.}\ \bibnamefont
  {Davies}} (\bibinfo {collaboration} {HPQCD Collaboration}),\ }\href {\doibase
  10.1103/PhysRevD.105.094506} {\bibfield  {journal} {\bibinfo  {journal}
  {Phys. Rev. D}\ }\textbf {\bibinfo {volume} {105}},\ \bibinfo {pages}
  {094506} (\bibinfo {year} {2022})},\ \Eprint
  {http://arxiv.org/abs/2105.11433} {arXiv:2105.11433 [hep-lat]} \BibitemShut
  {NoStop}%
\bibitem [{\citenamefont {Hu}\ \emph {et~al.}(2020)\citenamefont {Hu},
  \citenamefont {Jin},\ and\ \citenamefont {Xiao}}]{Hu:2019bdf}%
  \BibitemOpen
  \bibfield  {author} {\bibinfo {author} {\bibfnamefont {X.-Q.}\ \bibnamefont
  {Hu}}, \bibinfo {author} {\bibfnamefont {S.-P.}\ \bibnamefont {Jin}}, \ and\
  \bibinfo {author} {\bibfnamefont {Z.-J.}\ \bibnamefont {Xiao}},\ }\href
  {\doibase 10.1088/1674-1137/44/5/053102} {\bibfield  {journal} {\bibinfo
  {journal} {Chin. Phys. C}\ }\textbf {\bibinfo {volume} {44}},\ \bibinfo
  {pages} {053102} (\bibinfo {year} {2020})},\ \Eprint
  {http://arxiv.org/abs/1912.03981} {arXiv:1912.03981 [hep-ph]} \BibitemShut
  {NoStop}%
\bibitem [{\citenamefont {Zhang}\ \emph {et~al.}(2021)\citenamefont {Zhang},
  \citenamefont {Zhong}, \citenamefont {Fu}, \citenamefont {Cheng},\ and\
  \citenamefont {Wu}}]{Zhang:2021wnv}%
  \BibitemOpen
  \bibfield  {author} {\bibinfo {author} {\bibfnamefont {Y.}~\bibnamefont
  {Zhang}}, \bibinfo {author} {\bibfnamefont {T.}~\bibnamefont {Zhong}},
  \bibinfo {author} {\bibfnamefont {H.-B.}\ \bibnamefont {Fu}}, \bibinfo
  {author} {\bibfnamefont {W.}~\bibnamefont {Cheng}}, \ and\ \bibinfo {author}
  {\bibfnamefont {X.-G.}\ \bibnamefont {Wu}},\ }\href {\doibase
  10.1103/PhysRevD.103.114024} {\bibfield  {journal} {\bibinfo  {journal}
  {Phys. Rev. D}\ }\textbf {\bibinfo {volume} {103}},\ \bibinfo {pages}
  {114024} (\bibinfo {year} {2021})},\ \Eprint
  {http://arxiv.org/abs/2104.00180} {arXiv:2104.00180 [hep-ph]} \BibitemShut
  {NoStop}%
\bibitem [{\citenamefont {Bordone}\ \emph {et~al.}(2020)\citenamefont
  {Bordone}, \citenamefont {Gubernari}, \citenamefont {van Dyk},\ and\
  \citenamefont {Jung}}]{Bordone:2019guc}%
  \BibitemOpen
  \bibfield  {author} {\bibinfo {author} {\bibfnamefont {M.}~\bibnamefont
  {Bordone}}, \bibinfo {author} {\bibfnamefont {N.}~\bibnamefont {Gubernari}},
  \bibinfo {author} {\bibfnamefont {D.}~\bibnamefont {van Dyk}}, \ and\
  \bibinfo {author} {\bibfnamefont {M.}~\bibnamefont {Jung}},\ }\href {\doibase
  10.1140/epjc/s10052-020-7850-9} {\bibfield  {journal} {\bibinfo  {journal}
  {Eur. Phys. J. C}\ }\textbf {\bibinfo {volume} {80}},\ \bibinfo {pages} {347}
  (\bibinfo {year} {2020})},\ \Eprint {http://arxiv.org/abs/1912.09335}
  {arXiv:1912.09335 [hep-ph]} \BibitemShut {NoStop}%
\bibitem [{\citenamefont {Li}\ \emph {et~al.}(2009)\citenamefont {Li},
  \citenamefont {Lu},\ and\ \citenamefont {Wang}}]{Li:2009wq}%
  \BibitemOpen
  \bibfield  {author} {\bibinfo {author} {\bibfnamefont {R.-H.}\ \bibnamefont
  {Li}}, \bibinfo {author} {\bibfnamefont {C.-D.}\ \bibnamefont {Lu}}, \ and\
  \bibinfo {author} {\bibfnamefont {Y.-M.}\ \bibnamefont {Wang}},\ }\href
  {\doibase 10.1103/PhysRevD.80.014005} {\bibfield  {journal} {\bibinfo
  {journal} {Phys. Rev. D}\ }\textbf {\bibinfo {volume} {80}},\ \bibinfo
  {pages} {014005} (\bibinfo {year} {2009})},\ \Eprint
  {http://arxiv.org/abs/0905.3259} {arXiv:0905.3259 [hep-ph]} \BibitemShut
  {NoStop}%
\bibitem [{\citenamefont {Blasi}\ \emph {et~al.}(1994)\citenamefont {Blasi},
  \citenamefont {Colangelo}, \citenamefont {Nardulli},\ and\ \citenamefont
  {Paver}}]{Blasi:1993fi}%
  \BibitemOpen
  \bibfield  {author} {\bibinfo {author} {\bibfnamefont {P.}~\bibnamefont
  {Blasi}}, \bibinfo {author} {\bibfnamefont {P.}~\bibnamefont {Colangelo}},
  \bibinfo {author} {\bibfnamefont {G.}~\bibnamefont {Nardulli}}, \ and\
  \bibinfo {author} {\bibfnamefont {N.}~\bibnamefont {Paver}},\ }\href
  {\doibase 10.1103/PhysRevD.49.238} {\bibfield  {journal} {\bibinfo  {journal}
  {Phys. Rev. D}\ }\textbf {\bibinfo {volume} {49}},\ \bibinfo {pages} {238}
  (\bibinfo {year} {1994})},\ \Eprint {http://arxiv.org/abs/hep-ph/9307290}
  {arXiv:hep-ph/9307290} \BibitemShut {NoStop}%
\bibitem [{\citenamefont {Verma}(2012)}]{Verma:2011yw}%
  \BibitemOpen
  \bibfield  {author} {\bibinfo {author} {\bibfnamefont {R.~C.}\ \bibnamefont
  {Verma}},\ }\href {\doibase 10.1088/0954-3899/39/2/025005} {\bibfield
  {journal} {\bibinfo  {journal} {J. Phys. G}\ }\textbf {\bibinfo {volume}
  {39}},\ \bibinfo {pages} {025005} (\bibinfo {year} {2012})},\ \Eprint
  {http://arxiv.org/abs/1103.2973} {arXiv:1103.2973 [hep-ph]} \BibitemShut
  {NoStop}%
\bibitem [{\citenamefont {Kang}\ \emph {et~al.}(2018)\citenamefont {Kang},
  \citenamefont {Luo}, \citenamefont {Zhang}, \citenamefont {Dai},\ and\
  \citenamefont {Wang}}]{Kang:2018jzg}%
  \BibitemOpen
  \bibfield  {author} {\bibinfo {author} {\bibfnamefont {X.-W.}\ \bibnamefont
  {Kang}}, \bibinfo {author} {\bibfnamefont {T.}~\bibnamefont {Luo}}, \bibinfo
  {author} {\bibfnamefont {Y.}~\bibnamefont {Zhang}}, \bibinfo {author}
  {\bibfnamefont {L.-Y.}\ \bibnamefont {Dai}}, \ and\ \bibinfo {author}
  {\bibfnamefont {C.}~\bibnamefont {Wang}},\ }\href {\doibase
  10.1140/epjc/s10052-018-6385-9} {\bibfield  {journal} {\bibinfo  {journal}
  {Eur. Phys. J. C}\ }\textbf {\bibinfo {volume} {78}},\ \bibinfo {pages} {909}
  (\bibinfo {year} {2018})},\ \Eprint {http://arxiv.org/abs/1808.02432}
  {arXiv:1808.02432 [hep-ph]} \BibitemShut {NoStop}%
\bibitem [{\citenamefont {Li}\ \emph {et~al.}(2010)\citenamefont {Li},
  \citenamefont {Shao},\ and\ \citenamefont {Wang}}]{Li:2010bb}%
  \BibitemOpen
  \bibfield  {author} {\bibinfo {author} {\bibfnamefont {G.}~\bibnamefont
  {Li}}, \bibinfo {author} {\bibfnamefont {F.-l.}\ \bibnamefont {Shao}}, \ and\
  \bibinfo {author} {\bibfnamefont {W.}~\bibnamefont {Wang}},\ }\href {\doibase
  10.1103/PhysRevD.82.094031} {\bibfield  {journal} {\bibinfo  {journal} {Phys.
  Rev. D}\ }\textbf {\bibinfo {volume} {82}},\ \bibinfo {pages} {094031}
  (\bibinfo {year} {2010})},\ \Eprint {http://arxiv.org/abs/1008.3696}
  {arXiv:1008.3696 [hep-ph]} \BibitemShut {NoStop}%
\bibitem [{\citenamefont {Faustov}\ and\ \citenamefont
  {Galkin}(2013)}]{Faustov:2012mt}%
  \BibitemOpen
  \bibfield  {author} {\bibinfo {author} {\bibfnamefont {R.~N.}\ \bibnamefont
  {Faustov}}\ and\ \bibinfo {author} {\bibfnamefont {V.~O.}\ \bibnamefont
  {Galkin}},\ }\href {\doibase 10.1103/PhysRevD.87.034033} {\bibfield
  {journal} {\bibinfo  {journal} {Phys. Rev. D}\ }\textbf {\bibinfo {volume}
  {87}},\ \bibinfo {pages} {034033} (\bibinfo {year} {2013})},\ \Eprint
  {http://arxiv.org/abs/1212.3167} {arXiv:1212.3167 [hep-ph]} \BibitemShut
  {NoStop}%
\bibitem [{\citenamefont {Chen}\ \emph {et~al.}(2012)\citenamefont {Chen},
  \citenamefont {Fu}, \citenamefont {Kim},\ and\ \citenamefont
  {Wang}}]{Chen:2011ut}%
  \BibitemOpen
  \bibfield  {author} {\bibinfo {author} {\bibfnamefont {X.~J.}\ \bibnamefont
  {Chen}}, \bibinfo {author} {\bibfnamefont {H.~F.}\ \bibnamefont {Fu}},
  \bibinfo {author} {\bibfnamefont {C.~S.}\ \bibnamefont {Kim}}, \ and\
  \bibinfo {author} {\bibfnamefont {G.~L.}\ \bibnamefont {Wang}},\ }\href
  {\doibase 10.1088/0954-3899/39/4/045002} {\bibfield  {journal} {\bibinfo
  {journal} {J. Phys. G}\ }\textbf {\bibinfo {volume} {39}},\ \bibinfo {pages}
  {045002} (\bibinfo {year} {2012})},\ \Eprint {http://arxiv.org/abs/1106.3003}
  {arXiv:1106.3003 [hep-ph]} \BibitemShut {NoStop}%
\bibitem [{\citenamefont {Zhao}\ \emph {et~al.}(2007)\citenamefont {Zhao},
  \citenamefont {Liu},\ and\ \citenamefont {Li}}]{Zhao:2006at}%
  \BibitemOpen
  \bibfield  {author} {\bibinfo {author} {\bibfnamefont {S.-M.}\ \bibnamefont
  {Zhao}}, \bibinfo {author} {\bibfnamefont {X.}~\bibnamefont {Liu}}, \ and\
  \bibinfo {author} {\bibfnamefont {S.-J.}\ \bibnamefont {Li}},\ }\href
  {\doibase 10.1140/epjc/s10052-007-0322-7} {\bibfield  {journal} {\bibinfo
  {journal} {Eur. Phys. J. C}\ }\textbf {\bibinfo {volume} {51}},\ \bibinfo
  {pages} {601} (\bibinfo {year} {2007})},\ \Eprint
  {http://arxiv.org/abs/hep-ph/0612008} {arXiv:hep-ph/0612008} \BibitemShut
  {NoStop}%
\bibitem [{\citenamefont {Efimov}\ and\ \citenamefont
  {Ivanov}(1993)}]{Efimov:1993}%
  \BibitemOpen
  \bibfield  {author} {\bibinfo {author} {\bibfnamefont {G.~V.}\ \bibnamefont
  {Efimov}}\ and\ \bibinfo {author} {\bibfnamefont {M.~A.}\ \bibnamefont
  {Ivanov}},\ }\href@noop {} {\emph {\bibinfo {title} {{The Quark confinement
  model of hadrons}}}}\ (\bibinfo  {publisher} {IOP},\ \bibinfo {address}
  {Bristol},\ \bibinfo {year} {1993})\BibitemShut {NoStop}%
\bibitem [{\citenamefont {Branz}\ \emph {et~al.}(2010)\citenamefont {Branz},
  \citenamefont {Faessler}, \citenamefont {Gutsche}, \citenamefont {Ivanov},
  \citenamefont {Korner},\ and\ \citenamefont {Lyubovitskij}}]{Branz:2009cd}%
  \BibitemOpen
  \bibfield  {author} {\bibinfo {author} {\bibfnamefont {T.}~\bibnamefont
  {Branz}}, \bibinfo {author} {\bibfnamefont {A.}~\bibnamefont {Faessler}},
  \bibinfo {author} {\bibfnamefont {T.}~\bibnamefont {Gutsche}}, \bibinfo
  {author} {\bibfnamefont {M.~A.}\ \bibnamefont {Ivanov}}, \bibinfo {author}
  {\bibfnamefont {J.~G.}\ \bibnamefont {Korner}}, \ and\ \bibinfo {author}
  {\bibfnamefont {V.~E.}\ \bibnamefont {Lyubovitskij}},\ }\href {\doibase
  10.1103/PhysRevD.81.034010} {\bibfield  {journal} {\bibinfo  {journal} {Phys.
  Rev. D}\ }\textbf {\bibinfo {volume} {81}},\ \bibinfo {pages} {034010}
  (\bibinfo {year} {2010})},\ \Eprint {http://arxiv.org/abs/0912.3710}
  {arXiv:0912.3710 [hep-ph]} \BibitemShut {NoStop}%
\bibitem [{\citenamefont {Ivanov}\ \emph {et~al.}(2012)\citenamefont {Ivanov},
  \citenamefont {Korner}, \citenamefont {Kovalenko}, \citenamefont
  {Santorelli},\ and\ \citenamefont {Saidullaeva}}]{Ivanov:2011aa}%
  \BibitemOpen
  \bibfield  {author} {\bibinfo {author} {\bibfnamefont {M.~A.}\ \bibnamefont
  {Ivanov}}, \bibinfo {author} {\bibfnamefont {J.~G.}\ \bibnamefont {Korner}},
  \bibinfo {author} {\bibfnamefont {S.~G.}\ \bibnamefont {Kovalenko}}, \bibinfo
  {author} {\bibfnamefont {P.}~\bibnamefont {Santorelli}}, \ and\ \bibinfo
  {author} {\bibfnamefont {G.~G.}\ \bibnamefont {Saidullaeva}},\ }\href
  {\doibase 10.1103/PhysRevD.85.034004} {\bibfield  {journal} {\bibinfo
  {journal} {Phys. Rev. D}\ }\textbf {\bibinfo {volume} {85}},\ \bibinfo
  {pages} {034004} (\bibinfo {year} {2012})},\ \Eprint
  {http://arxiv.org/abs/1112.3536} {arXiv:1112.3536 [hep-ph]} \BibitemShut
  {NoStop}%
\bibitem [{\citenamefont {Gutsche}\ \emph {et~al.}(2012)\citenamefont
  {Gutsche}, \citenamefont {Ivanov}, \citenamefont {Korner}, \citenamefont
  {Lyubovitskij},\ and\ \citenamefont {Santorelli}}]{Gutsche:2012ze}%
  \BibitemOpen
  \bibfield  {author} {\bibinfo {author} {\bibfnamefont {T.}~\bibnamefont
  {Gutsche}}, \bibinfo {author} {\bibfnamefont {M.~A.}\ \bibnamefont {Ivanov}},
  \bibinfo {author} {\bibfnamefont {J.~G.}\ \bibnamefont {Korner}}, \bibinfo
  {author} {\bibfnamefont {V.~E.}\ \bibnamefont {Lyubovitskij}}, \ and\
  \bibinfo {author} {\bibfnamefont {P.}~\bibnamefont {Santorelli}},\ }\href
  {\doibase 10.1103/PhysRevD.86.074013} {\bibfield  {journal} {\bibinfo
  {journal} {Phys. Rev. D}\ }\textbf {\bibinfo {volume} {86}},\ \bibinfo
  {pages} {074013} (\bibinfo {year} {2012})},\ \Eprint
  {http://arxiv.org/abs/1207.7052} {arXiv:1207.7052 [hep-ph]} \BibitemShut
  {NoStop}%
\bibitem [{\citenamefont {Soni}\ \emph {et~al.}(2022)\citenamefont {Soni},
  \citenamefont {Issadykov}, \citenamefont {Gadaria}, \citenamefont {Patel},\
  and\ \citenamefont {Pandya}}]{Soni:2020bvu}%
  \BibitemOpen
  \bibfield  {author} {\bibinfo {author} {\bibfnamefont {N.~R.}\ \bibnamefont
  {Soni}}, \bibinfo {author} {\bibfnamefont {A.}~\bibnamefont {Issadykov}},
  \bibinfo {author} {\bibfnamefont {A.~N.}\ \bibnamefont {Gadaria}}, \bibinfo
  {author} {\bibfnamefont {J.~J.}\ \bibnamefont {Patel}}, \ and\ \bibinfo
  {author} {\bibfnamefont {J.~N.}\ \bibnamefont {Pandya}},\ }\href {\doibase
  10.1140/epja/s10050-022-00685-y} {\bibfield  {journal} {\bibinfo  {journal}
  {Eur. Phys. J. A}\ }\textbf {\bibinfo {volume} {58}},\ \bibinfo {pages} {39}
  (\bibinfo {year} {2022})},\ \Eprint {http://arxiv.org/abs/2008.07202}
  {arXiv:2008.07202 [hep-ph]} \BibitemShut {NoStop}%
\bibitem [{\citenamefont {Soni}\ and\ \citenamefont
  {Pandya}(2017)}]{Soni:2017eug}%
  \BibitemOpen
  \bibfield  {author} {\bibinfo {author} {\bibfnamefont {N.~R.}\ \bibnamefont
  {Soni}}\ and\ \bibinfo {author} {\bibfnamefont {J.~N.}\ \bibnamefont
  {Pandya}},\ }\href {\doibase 10.1103/PhysRevD.96.016017} {\bibfield
  {journal} {\bibinfo  {journal} {Phys. Rev. D}\ }\textbf {\bibinfo {volume}
  {96}},\ \bibinfo {pages} {016017} (\bibinfo {year} {2017})},\ \bibinfo {note}
  {[Erratum: Phys.Rev.D 99, 059901 (2019)]},\ \Eprint
  {http://arxiv.org/abs/1706.01190} {arXiv:1706.01190 [hep-ph]} \BibitemShut
  {NoStop}%
\bibitem [{\citenamefont {Soni}\ \emph {et~al.}(2018)\citenamefont {Soni},
  \citenamefont {Ivanov}, \citenamefont {K\"orner}, \citenamefont {Pandya},
  \citenamefont {Santorelli},\ and\ \citenamefont {Tran}}]{Soni:2018adu}%
  \BibitemOpen
  \bibfield  {author} {\bibinfo {author} {\bibfnamefont {N.~R.}\ \bibnamefont
  {Soni}}, \bibinfo {author} {\bibfnamefont {M.~A.}\ \bibnamefont {Ivanov}},
  \bibinfo {author} {\bibfnamefont {J.~G.}\ \bibnamefont {K\"orner}}, \bibinfo
  {author} {\bibfnamefont {J.~N.}\ \bibnamefont {Pandya}}, \bibinfo {author}
  {\bibfnamefont {P.}~\bibnamefont {Santorelli}}, \ and\ \bibinfo {author}
  {\bibfnamefont {C.~T.}\ \bibnamefont {Tran}},\ }\href {\doibase
  10.1103/PhysRevD.98.114031} {\bibfield  {journal} {\bibinfo  {journal} {Phys.
  Rev. D}\ }\textbf {\bibinfo {volume} {98}},\ \bibinfo {pages} {114031}
  (\bibinfo {year} {2018})},\ \Eprint {http://arxiv.org/abs/1810.11907}
  {arXiv:1810.11907 [hep-ph]} \BibitemShut {NoStop}%
\bibitem [{\citenamefont {Ivanov}\ \emph {et~al.}(2019)\citenamefont {Ivanov},
  \citenamefont {K\"orner}, \citenamefont {Pandya}, \citenamefont {Santorelli},
  \citenamefont {Soni},\ and\ \citenamefont {Tran}}]{Ivanov:2019nqd}%
  \BibitemOpen
  \bibfield  {author} {\bibinfo {author} {\bibfnamefont {M.~A.}\ \bibnamefont
  {Ivanov}}, \bibinfo {author} {\bibfnamefont {J.~G.}\ \bibnamefont
  {K\"orner}}, \bibinfo {author} {\bibfnamefont {J.~N.}\ \bibnamefont
  {Pandya}}, \bibinfo {author} {\bibfnamefont {P.}~\bibnamefont {Santorelli}},
  \bibinfo {author} {\bibfnamefont {N.~R.}\ \bibnamefont {Soni}}, \ and\
  \bibinfo {author} {\bibfnamefont {C.-T.}\ \bibnamefont {Tran}},\ }\href
  {\doibase 10.1007/s11467-019-0908-1} {\bibfield  {journal} {\bibinfo
  {journal} {Front. Phys. (Beijing)}\ }\textbf {\bibinfo {volume} {14}},\
  \bibinfo {pages} {64401} (\bibinfo {year} {2019})},\ \Eprint
  {http://arxiv.org/abs/1904.07740} {arXiv:1904.07740 [hep-ph]} \BibitemShut
  {NoStop}%
\bibitem [{\citenamefont {Soni}\ \emph {et~al.}(2020)\citenamefont {Soni},
  \citenamefont {Gadaria}, \citenamefont {Patel},\ and\ \citenamefont
  {Pandya}}]{Soni:2020sgn}%
  \BibitemOpen
  \bibfield  {author} {\bibinfo {author} {\bibfnamefont {N.~R.}\ \bibnamefont
  {Soni}}, \bibinfo {author} {\bibfnamefont {A.~N.}\ \bibnamefont {Gadaria}},
  \bibinfo {author} {\bibfnamefont {J.~J.}\ \bibnamefont {Patel}}, \ and\
  \bibinfo {author} {\bibfnamefont {J.~N.}\ \bibnamefont {Pandya}},\ }\href
  {\doibase 10.1103/PhysRevD.102.016013} {\bibfield  {journal} {\bibinfo
  {journal} {Phys. Rev. D}\ }\textbf {\bibinfo {volume} {102}},\ \bibinfo
  {pages} {016013} (\bibinfo {year} {2020})},\ \Eprint
  {http://arxiv.org/abs/2001.10195} {arXiv:2001.10195 [hep-ph]} \BibitemShut
  {NoStop}%
\bibitem [{\citenamefont {Soni}\ and\ \citenamefont
  {Pandya}(2019{\natexlab{a}})}]{Soni:2019huk}%
  \BibitemOpen
  \bibfield  {author} {\bibinfo {author} {\bibfnamefont {N.~R.}\ \bibnamefont
  {Soni}}\ and\ \bibinfo {author} {\bibfnamefont {J.~N.}\ \bibnamefont
  {Pandya}},\ }\href {\doibase 10.1051/epjconf/201920206010} {\bibfield
  {journal} {\bibinfo  {journal} {EPJ Web Conf.}\ }\textbf {\bibinfo {volume}
  {202}},\ \bibinfo {pages} {06010} (\bibinfo {year}
  {2019}{\natexlab{a}})}\BibitemShut {NoStop}%
\bibitem [{\citenamefont {Soni}\ and\ \citenamefont
  {Pandya}(2019{\natexlab{b}})}]{Soni:2019qjs}%
  \BibitemOpen
  \bibfield  {author} {\bibinfo {author} {\bibfnamefont {N.~R.}\ \bibnamefont
  {Soni}}\ and\ \bibinfo {author} {\bibfnamefont {J.~N.}\ \bibnamefont
  {Pandya}},\ }\href {\doibase 10.1007/978-3-030-29622-3_16} {\bibfield
  {journal} {\bibinfo  {journal} {Springer Proc. Phys.}\ }\textbf {\bibinfo
  {volume} {234}},\ \bibinfo {pages} {115} (\bibinfo {year}
  {2019}{\natexlab{b}})}\BibitemShut {NoStop}%
\bibitem [{\citenamefont {Soni}\ and\ \citenamefont
  {Pandya}(2021)}]{Soni:2021rem}%
  \BibitemOpen
  \bibfield  {author} {\bibinfo {author} {\bibfnamefont {N.~R.}\ \bibnamefont
  {Soni}}\ and\ \bibinfo {author} {\bibfnamefont {J.~N.}\ \bibnamefont
  {Pandya}},\ }\href {\doibase 10.1007/978-981-33-4408-2_13} {\bibfield
  {journal} {\bibinfo  {journal} {Springer Proc. Phys.}\ }\textbf {\bibinfo
  {volume} {261}},\ \bibinfo {pages} {85} (\bibinfo {year} {2021})}\BibitemShut
  {NoStop}%
\bibitem [{\citenamefont {Dubnicka}\ \emph {et~al.}(2017)\citenamefont
  {Dubnicka}, \citenamefont {Dubnickova}, \citenamefont {Issadykov},
  \citenamefont {Ivanov},\ and\ \citenamefont {Liptaj}}]{Dubnicka:2017job}%
  \BibitemOpen
  \bibfield  {author} {\bibinfo {author} {\bibfnamefont {S.}~\bibnamefont
  {Dubnicka}}, \bibinfo {author} {\bibfnamefont {A.~Z.}\ \bibnamefont
  {Dubnickova}}, \bibinfo {author} {\bibfnamefont {A.}~\bibnamefont
  {Issadykov}}, \bibinfo {author} {\bibfnamefont {M.~A.}\ \bibnamefont
  {Ivanov}}, \ and\ \bibinfo {author} {\bibfnamefont {A.}~\bibnamefont
  {Liptaj}},\ }\href {\doibase 10.1103/PhysRevD.96.076017} {\bibfield
  {journal} {\bibinfo  {journal} {Phys. Rev. D}\ }\textbf {\bibinfo {volume}
  {96}},\ \bibinfo {pages} {076017} (\bibinfo {year} {2017})},\ \Eprint
  {http://arxiv.org/abs/1708.09607} {arXiv:1708.09607 [hep-ph]} \BibitemShut
  {NoStop}%
\bibitem [{\citenamefont {Issadykov}(2018)}]{Issadykov:2018jvt}%
  \BibitemOpen
  \bibfield  {author} {\bibinfo {author} {\bibfnamefont {A.}~\bibnamefont
  {Issadykov}},\ }\href {\doibase 10.1051/epjconf/201817709007} {\bibfield
  {journal} {\bibinfo  {journal} {EPJ Web Conf.}\ }\textbf {\bibinfo {volume}
  {177}},\ \bibinfo {pages} {09007} (\bibinfo {year} {2018})}\BibitemShut
  {NoStop}%
\bibitem [{\citenamefont {Issadykov}\ \emph {et~al.}(2017)\citenamefont
  {Issadykov}, \citenamefont {Ivanov},\ and\ \citenamefont
  {Nurbakova}}]{Issadykov:2017wlb}%
  \BibitemOpen
  \bibfield  {author} {\bibinfo {author} {\bibfnamefont {A.}~\bibnamefont
  {Issadykov}}, \bibinfo {author} {\bibfnamefont {M.~A.}\ \bibnamefont
  {Ivanov}}, \ and\ \bibinfo {author} {\bibfnamefont {G.}~\bibnamefont
  {Nurbakova}},\ }\href {\doibase 10.1051/epjconf/201715803002} {\bibfield
  {journal} {\bibinfo  {journal} {EPJ Web Conf.}\ }\textbf {\bibinfo {volume}
  {158}},\ \bibinfo {pages} {03002} (\bibinfo {year} {2017})},\ \Eprint
  {http://arxiv.org/abs/1907.13210} {arXiv:1907.13210 [hep-ph]} \BibitemShut
  {NoStop}%
\bibitem [{\citenamefont {Ivanov}\ \emph {et~al.}(2015)\citenamefont {Ivanov},
  \citenamefont {K\"orner},\ and\ \citenamefont {Tran}}]{Ivanov:2015tru}%
  \BibitemOpen
  \bibfield  {author} {\bibinfo {author} {\bibfnamefont {M.~A.}\ \bibnamefont
  {Ivanov}}, \bibinfo {author} {\bibfnamefont {J.~G.}\ \bibnamefont
  {K\"orner}}, \ and\ \bibinfo {author} {\bibfnamefont {C.~T.}\ \bibnamefont
  {Tran}},\ }\href {\doibase 10.1103/PhysRevD.92.114022} {\bibfield  {journal}
  {\bibinfo  {journal} {Phys. Rev. D}\ }\textbf {\bibinfo {volume} {92}},\
  \bibinfo {pages} {114022} (\bibinfo {year} {2015})},\ \Eprint
  {http://arxiv.org/abs/1508.02678} {arXiv:1508.02678 [hep-ph]} \BibitemShut
  {NoStop}%
\bibitem [{\citenamefont {Ivanov}\ \emph {et~al.}(2016)\citenamefont {Ivanov},
  \citenamefont {K\"orner},\ and\ \citenamefont {Tran}}]{Ivanov:2016qtw}%
  \BibitemOpen
  \bibfield  {author} {\bibinfo {author} {\bibfnamefont {M.~A.}\ \bibnamefont
  {Ivanov}}, \bibinfo {author} {\bibfnamefont {J.~G.}\ \bibnamefont
  {K\"orner}}, \ and\ \bibinfo {author} {\bibfnamefont {C.-T.}\ \bibnamefont
  {Tran}},\ }\href {\doibase 10.1103/PhysRevD.94.094028} {\bibfield  {journal}
  {\bibinfo  {journal} {Phys. Rev. D}\ }\textbf {\bibinfo {volume} {94}},\
  \bibinfo {pages} {094028} (\bibinfo {year} {2016})},\ \Eprint
  {http://arxiv.org/abs/1607.02932} {arXiv:1607.02932 [hep-ph]} \BibitemShut
  {NoStop}%
\bibitem [{\citenamefont {Ivanov}\ \emph {et~al.}(2017)\citenamefont {Ivanov},
  \citenamefont {K\"orner},\ and\ \citenamefont {Tran}}]{Ivanov:2017mrj}%
  \BibitemOpen
  \bibfield  {author} {\bibinfo {author} {\bibfnamefont {M.~A.}\ \bibnamefont
  {Ivanov}}, \bibinfo {author} {\bibfnamefont {J.~G.}\ \bibnamefont
  {K\"orner}}, \ and\ \bibinfo {author} {\bibfnamefont {C.-T.}\ \bibnamefont
  {Tran}},\ }\href {\doibase 10.1103/PhysRevD.95.036021} {\bibfield  {journal}
  {\bibinfo  {journal} {Phys. Rev. D}\ }\textbf {\bibinfo {volume} {95}},\
  \bibinfo {pages} {036021} (\bibinfo {year} {2017})},\ \Eprint
  {http://arxiv.org/abs/1701.02937} {arXiv:1701.02937 [hep-ph]} \BibitemShut
  {NoStop}%
\bibitem [{\citenamefont {Weinberg}(1963)}]{Weinberg:1962}%
  \BibitemOpen
  \bibfield  {author} {\bibinfo {author} {\bibfnamefont {S.}~\bibnamefont
  {Weinberg}},\ }\href {\doibase 10.1103/PhysRev.130.776} {\bibfield  {journal}
  {\bibinfo  {journal} {Phys. Rev.}\ }\textbf {\bibinfo {volume} {130}},\
  \bibinfo {pages} {776} (\bibinfo {year} {1963})}\BibitemShut {NoStop}%
\bibitem [{\citenamefont {Salam}(1962)}]{Salam:1962}%
  \BibitemOpen
  \bibfield  {author} {\bibinfo {author} {\bibfnamefont {A.}~\bibnamefont
  {Salam}},\ }\href {\doibase 10.1007/BF02733330} {\bibfield  {journal}
  {\bibinfo  {journal} {Nuovo Cim.}\ }\textbf {\bibinfo {volume} {25}},\
  \bibinfo {pages} {224} (\bibinfo {year} {1962})}\BibitemShut {NoStop}%
\bibitem [{\citenamefont {Ivanov}\ and\ \citenamefont
  {Lyubovitskij}(1997)}]{Ivanov:1997ug}%
  \BibitemOpen
  \bibfield  {author} {\bibinfo {author} {\bibfnamefont {M.~A.}\ \bibnamefont
  {Ivanov}}\ and\ \bibinfo {author} {\bibfnamefont {V.~E.}\ \bibnamefont
  {Lyubovitskij}},\ }\href {\doibase 10.1016/S0370-2693(97)00776-4} {\bibfield
  {journal} {\bibinfo  {journal} {Phys. Lett. B}\ }\textbf {\bibinfo {volume}
  {408}},\ \bibinfo {pages} {435} (\bibinfo {year} {1997})},\ \Eprint
  {http://arxiv.org/abs/hep-ph/9705423} {arXiv:hep-ph/9705423} \BibitemShut
  {NoStop}%
\bibitem [{\citenamefont {Faessler}\ \emph {et~al.}(2003)\citenamefont
  {Faessler}, \citenamefont {Gutsche}, \citenamefont {Ivanov}, \citenamefont
  {Lyubovitskij},\ and\ \citenamefont {Wang}}]{Faessler:2003yf}%
  \BibitemOpen
  \bibfield  {author} {\bibinfo {author} {\bibfnamefont {A.}~\bibnamefont
  {Faessler}}, \bibinfo {author} {\bibfnamefont {T.}~\bibnamefont {Gutsche}},
  \bibinfo {author} {\bibfnamefont {M.~A.}\ \bibnamefont {Ivanov}}, \bibinfo
  {author} {\bibfnamefont {V.~E.}\ \bibnamefont {Lyubovitskij}}, \ and\
  \bibinfo {author} {\bibfnamefont {P.}~\bibnamefont {Wang}},\ }\href {\doibase
  10.1103/PhysRevD.68.014011} {\bibfield  {journal} {\bibinfo  {journal} {Phys.
  Rev. D}\ }\textbf {\bibinfo {volume} {68}},\ \bibinfo {pages} {014011}
  (\bibinfo {year} {2003})},\ \Eprint {http://arxiv.org/abs/hep-ph/0304031}
  {arXiv:hep-ph/0304031} \BibitemShut {NoStop}%
\bibitem [{\citenamefont {Zyla}\ \emph {et~al.}(2020)\citenamefont {Zyla} \emph
  {et~al.}}]{ParticleDataGroup:2020ssz}%
  \BibitemOpen
  \bibfield  {author} {\bibinfo {author} {\bibfnamefont {P.~A.}\ \bibnamefont
  {Zyla}} \emph {et~al.} (\bibinfo {collaboration} {Particle Data Group}),\
  }\href {\doibase 10.1093/ptep/ptaa104} {\bibfield  {journal} {\bibinfo
  {journal} {PTEP}\ }\textbf {\bibinfo {volume} {2020}},\ \bibinfo {pages}
  {083C01} (\bibinfo {year} {2020})}\BibitemShut {NoStop}%
\bibitem [{\citenamefont {Alberti}\ \emph {et~al.}(2015)\citenamefont
  {Alberti}, \citenamefont {Gambino}, \citenamefont {Healey},\ and\
  \citenamefont {Nandi}}]{Alberti:2014yda}%
  \BibitemOpen
  \bibfield  {author} {\bibinfo {author} {\bibfnamefont {A.}~\bibnamefont
  {Alberti}}, \bibinfo {author} {\bibfnamefont {P.}~\bibnamefont {Gambino}},
  \bibinfo {author} {\bibfnamefont {K.~J.}\ \bibnamefont {Healey}}, \ and\
  \bibinfo {author} {\bibfnamefont {S.}~\bibnamefont {Nandi}},\ }\href
  {\doibase 10.1103/PhysRevLett.114.061802} {\bibfield  {journal} {\bibinfo
  {journal} {Phys. Rev. Lett.}\ }\textbf {\bibinfo {volume} {114}},\ \bibinfo
  {pages} {061802} (\bibinfo {year} {2015})},\ \Eprint
  {http://arxiv.org/abs/1411.6560} {arXiv:1411.6560 [hep-ph]} \BibitemShut
  {NoStop}%
\bibitem [{\citenamefont {Aaij}\ \emph
  {et~al.}(2020{\natexlab{b}})\citenamefont {Aaij} \emph
  {et~al.}}]{LHCb:2020hpv}%
  \BibitemOpen
  \bibfield  {author} {\bibinfo {author} {\bibfnamefont {R.}~\bibnamefont
  {Aaij}} \emph {et~al.} (\bibinfo {collaboration} {LHCb Collaboration}),\
  }\href {\doibase 10.1007/JHEP12(2020)144} {\bibfield  {journal} {\bibinfo
  {journal} {JHEP}\ }\textbf {\bibinfo {volume} {12}},\ \bibinfo {pages} {144}
  (\bibinfo {year} {2020}{\natexlab{b}})},\ \Eprint
  {http://arxiv.org/abs/2003.08453} {arXiv:2003.08453 [hep-ex]} \BibitemShut
  {NoStop}%
\bibitem [{\citenamefont {Wirbel}\ \emph {et~al.}(1985)\citenamefont {Wirbel},
  \citenamefont {Stech},\ and\ \citenamefont {Bauer}}]{Wirbel:1985ji}%
  \BibitemOpen
  \bibfield  {author} {\bibinfo {author} {\bibfnamefont {M.}~\bibnamefont
  {Wirbel}}, \bibinfo {author} {\bibfnamefont {B.}~\bibnamefont {Stech}}, \
  and\ \bibinfo {author} {\bibfnamefont {M.}~\bibnamefont {Bauer}},\ }\href
  {\doibase 10.1007/BF01560299} {\bibfield  {journal} {\bibinfo  {journal} {Z.
  Phys. C}\ }\textbf {\bibinfo {volume} {29}},\ \bibinfo {pages} {637}
  (\bibinfo {year} {1985})}\BibitemShut {NoStop}%
\bibitem [{\citenamefont {Issadykov}\ and\ \citenamefont
  {Ivanov}(2018)}]{Issadykov:2018myx}%
  \BibitemOpen
  \bibfield  {author} {\bibinfo {author} {\bibfnamefont {A.}~\bibnamefont
  {Issadykov}}\ and\ \bibinfo {author} {\bibfnamefont {M.~A.}\ \bibnamefont
  {Ivanov}},\ }\href {\doibase 10.1016/j.physletb.2018.06.056} {\bibfield
  {journal} {\bibinfo  {journal} {Phys. Lett. B}\ }\textbf {\bibinfo {volume}
  {783}},\ \bibinfo {pages} {178} (\bibinfo {year} {2018})},\ \Eprint
  {http://arxiv.org/abs/1804.00472} {arXiv:1804.00472 [hep-ph]} \BibitemShut
  {NoStop}%
\bibitem [{\citenamefont {Caria}\ \emph {et~al.}(2020)\citenamefont {Caria}
  \emph {et~al.}}]{Belle:2019rba}%
  \BibitemOpen
  \bibfield  {author} {\bibinfo {author} {\bibfnamefont {G.}~\bibnamefont
  {Caria}} \emph {et~al.} (\bibinfo {collaboration} {Belle Collaboration}),\
  }\href {\doibase 10.1103/PhysRevLett.124.161803} {\bibfield  {journal}
  {\bibinfo  {journal} {Phys. Rev. Lett.}\ }\textbf {\bibinfo {volume} {124}},\
  \bibinfo {pages} {161803} (\bibinfo {year} {2020})},\ \Eprint
  {http://arxiv.org/abs/1910.05864} {arXiv:1910.05864 [hep-ex]} \BibitemShut
  {NoStop}%
\end{thebibliography}%
\end{document}